\documentclass[12pt]{article}

\usepackage[dvips]{color}
\usepackage{amsmath}
\usepackage{graphicx} 
\usepackage[dvips]{color}
\usepackage{graphicx}
\usepackage{tabularx}
\usepackage{subfig}

\newcommand{\toolboxName}{{\ttfamily{BrainGeneExpressionAnalysis}}}
\newcommand{\toolboxNameAcronym}{BGEA}
\newcommand{\toolboxLink}{{\ttfamily{www.brainarchitecture.org/allen-atlas-brain-toolbox}}}
\newcommand{\figWidth}{6.5in}
\newcommand{\GAtlas}{G_{\mathrm{atlas}}}
\newcommand{\GSet}{G_{\mathrm{set}}}
\newcommand{\CAtlas}{C^{\mathrm{atlas}}}
\newcommand{\CSet}{C^{\mathrm{set}}}
\newcommand{\numDraws}{{\mathcal{N}}}
\begin{document}
\title{Computational neuroanatomy and co-expression 
 of genes in the adult mouse brain, analysis tools for the Allen Brain Atlas}
\date{}
\author{Pascal Grange$\;^1$, Michael Hawrylycz$\;^2$ and Partha P. Mitra$\;^1$ \\
 \multicolumn{1}{p{.9\textwidth}}{\normalsize{\centering\emph{$\;^1$ Cold Spring Harbor Laboratory,
 One Bungtown Road, Cold Spring Harbor, {\hbox{New York 11724}}, United States}}}\\
  \multicolumn{1}{p{.9\textwidth}}{\normalsize{\centering\emph{$\;^2$ Allen Institute for Brain Science, Seattle, Washington 98103, United States}}}}
\maketitle

\begin{abstract}
 We review quantitative methods and software developed to analyze
 genome-scale, brain-wide spatially-mapped gene-expression data. We
 expose new methods based on the underlying high-dimensional geometry
 of voxel space and gene space, and on simulations of the distribution
 of co-expression networks of a given size. We apply them to the Allen
 Atlas of the adult mouse brain, and to the co-expression network of a
 set of genes related to nicotine addiction retrieved from the NicSNP
 database. The computational methods are implemented in \toolboxName,
 a Matlab toolbox available for download.
 
\end{abstract}

\clearpage

\tableofcontents

\section{Introduction and background}

The mammalian brain is a structure of daunting complexity, whose study
started millenia ago and has been recently renewed by molecular
biology and computational imaging \cite{GeneToBrain}. The Allen Brain
Atlas, the first Web-based, genome-wide atlas of gene expression in
the adult mouse brain, was a large-scale experimental effort
\cite{AllenGenome,images,BrainAtlasInsights,
neufoAllen,AllenFiveYears,corrStructureAllen}. The
resulting dataset consists of co-registered {\emph{in situ}}
hybridization (ISH) image series for thousands of genes. It is now
available to neuroscientists world-wide, and has given rise to the
development of quantitative techniques and software for data
analysis. The present paper reviews recent developments that have been
applied to co-expression studies in the mouse brain and are publicly
available for use on the Web \cite{addiction} and on the desktop
\cite{toolboxManual}.\\

  On the other hand, lists of condition-related genes are now available 
from databases that pool results of different studies \cite{KARG, NicSNP}.
 As these studies employ different methods and result 
 in lists of hundreds of  genes, it is important to investigate 
 any possible order (or lack of it) in these lists. The Allen Brain Atlas
 provides ways to do this, by stuying brain-wide co-expression of genes,
 and by enabling to compare gene expression to classical neuroanatomy,
 in a genome-scale dataset based on a unified protocol.\\

 Advanced data exploration tools have already been
developed for the Allen Brain Atlas. NeuroBlast allows users to
explore the correlation structure between genes in the ABA. was
inspired by the Basic Local Alignment Research Tool \cite{BLAST},
which derives lists of similar genes to a given gene at the level of
sequences, and transposed the technique to the analysis of similarity
between patterns of gene expression in the brain \cite{NeuroBlast}.
The Anatomic Gene Expression Atlas \cite{AGEA} was launched in 2007.
 It is based on the spatial correlation of the atlas. 
 The user can explore three-dimensional correlation maps based on 
 correlations between voxels, computed using thousands of genes, and
 retrieve hierarchical data-driven parcellations of the brain.\\

  The Weighted Gene Co-Expression Network Analysis framework (WGCNA) has been
 used to isolate clusters of genes from correlations between 
 multiple microarray samples.
  In this approach the gene networks are typically constructed
 from the correlation coefficients of microarray data, from which graphs 
 are constructed and thresholded at a value chosen as as to satisfy certain 
 statistical criteria \cite{ZhangFramework,feeding}.  
 However, in the case of the Allen Brain Atlas, gene-expression data 
 are scaffolded by classical neuroanatomy, since ISH data
 are co-registered to the Allen Reference Atlas (ARA) \cite{AllenAtlas}. 
  The whole brain is voxelized, and the voxels
 are are annotated according to the brain region to which they 
 belong, which allows to compare the expression of sets of genes
 to brain regions (see Figure \ref{coExpressionSet} and \cite{markerGenes}).
 Hence we developed computational methods to:\\
 1. study the 
 whole range of co-expression values between pairs of genes;\\
 2. use the Allen Atlas as a probabilistic universe to estimate
  the distribution of co-expression networks;\\
 3. compare the expression patterns of highly co-expressed sets 
 of genes to classical neuroanatomy.\\

These methods are implemented in
\toolboxName$\;$(\toolboxNameAcronym), a Matlab toolbox downloadable
from {\ttfamily{www.brainarchitecture.org}}.  They are applied to a
set of 288 genes extracted from the NicSNP database, which have been
linked to nicotine dependence, based on the statistical significance
of allele frequency difference between cases and controls, and for
which mouse orthologs are found in the coronal Allen Atlas.\\


\section{Methods}
The spatial frequency of tissue-sectioning in the experimental
pipeline of the Allen Brain Atlas 
corresponds to slices with a thickness of 100 micrometers.
Each section was registered to a grid with a resolution of 100 microns
\cite{LauExploration,digitalAtlasing}. The induced three-dimensional grid was
sub-sampled to a resolution of 200 microns in order to increase the
overlap between different experiments. This
 procedure results in a partition of the mouse brain
 into $V=49,742$ cubic voxels. We focus on the co-registered
quantities obtained at a spatial resolution of 200 micrometers, for 
 several thousands of genes, after
 subsampling.\\

 In particular, the expression energy of each gene labelled $g$ in the
 Atlas was defined and computed \cite{AGEA} at each voxel
 labelled $v$ in the mouse brain:
\begin{equation}
E( v,g ) = \frac{\sum_{p\in v} M(p)I(p)} {\sum_{p\in v} 1 },
\label{expressionEnergy}
\end{equation}
where $p$ is a pixel index, and the denominator counts the pixels that are
contained in the voxel $v$ for 
 the ISH image series of gene $g$. The quantity $M(p)$ is a Boolean segmentation
mask that takes value $1$ at pixels classified as expressing the gene,
and 0 at other pixels. The quantity $I(p)$ is the grayscale value of
the pixels in ISH images. The present paper uses the voxel-by-gene
matrix of expression energies $E$ as the digitized version of the Allen
Brain Atlas. The expression energies of the genes in the full coronal and
sagittal atlas can be downloaded using the Web service provided by the
Allen Institute \cite{AllenWebServiceSite}.

\subsection{Brain-wide co-expression networks: graph properties}

\begin{figure}
\centering
\includegraphics[width=5in,keepaspectratio]{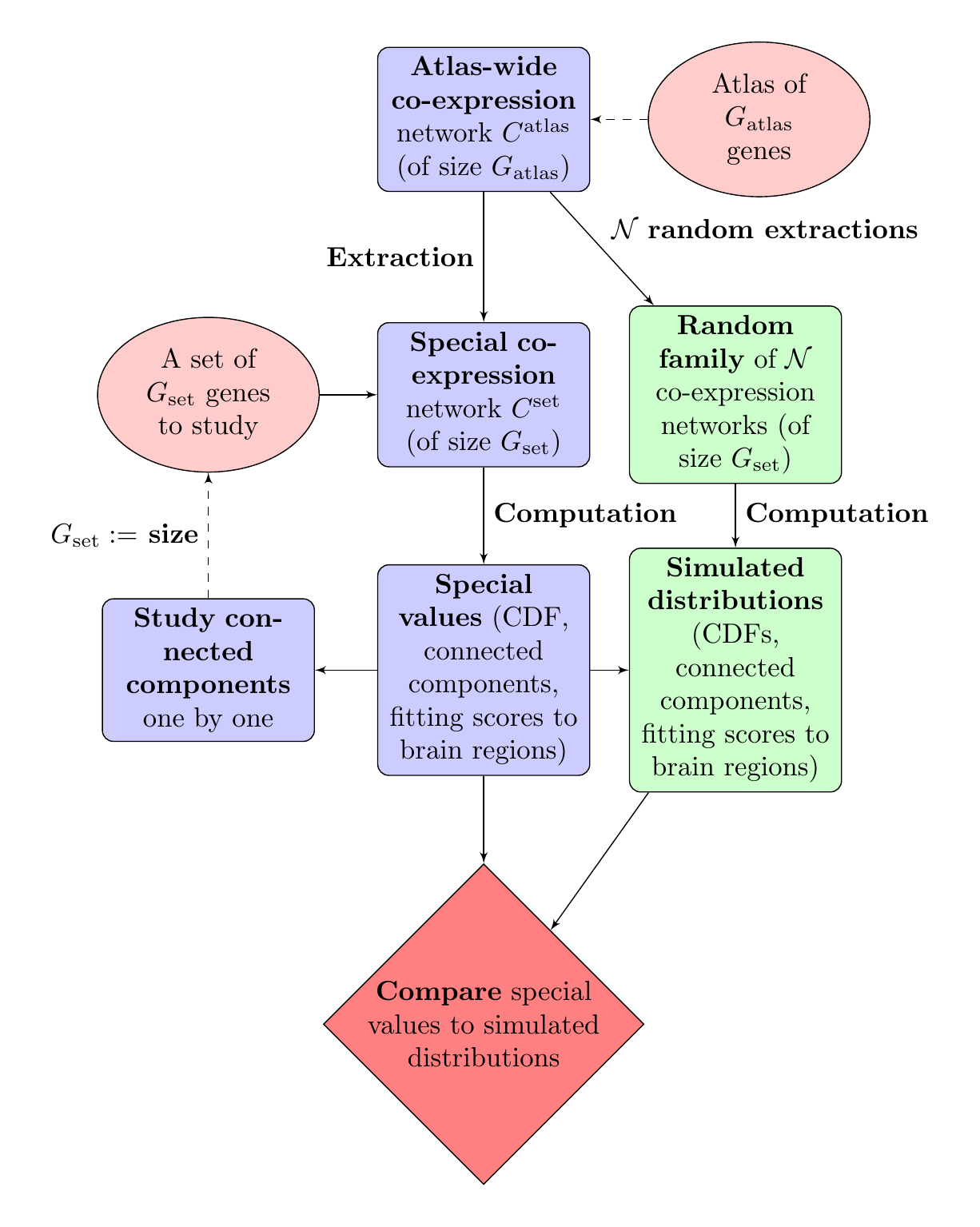}
\caption{The flowchart of computational analysis of the 
 collective neuroanatomical properties of a set of genes in the \toolboxName$\;$ toolbox.
 The steps marked as random extractions and computations are described in supplementary 
 materials.}
\label{flowChart}
\end{figure}

 The statistical study of brain-wide co-expression networks
 using \toolboxNameAcronym$\;$ 
  is summarized in the flowchart of Figure \ref{flowChart}. 
Detailed examples of the use of the software 
 are provided in toolbox manual \cite{toolboxManual}.\\

 The columns of the
matrix $E$ of expression energies of Equation \ref{expressionEnergy} are
naturally identified to vectors in a $V$-dimensional space (the voxel space).
  Given two genes, the two
corresponding columns of the matrix $E$ span a two-dimensional vector 
 of voxel space. The simplest geometric quantity to study for this system
 is the angle between the two vectors. As all the entries of
the matrix $E$ are positive by construction, this angle is between 0
and $\pi/2$. The angle between the two vectors is therefore completely
characterized by its cosine, which is readily expressed in terms of
expression energies. This cosine similarity, defined in Equation \ref{coExpr},
 for genes labelled $g$ and $g'$, is called the co-expression of genes 
 $g$ and $g'$.\\


\begin{equation}
{\mathrm{coExpr}}( g, g')= \sum_{v = 1}^V \frac{E(v,g) E(v,g')}{\sqrt{\sum_{u = 1}^VE(u,g )^2\sum_{w = 1}^VE(w,g )^2}}.
\label{coExpr}
\end{equation}
 The more co-expressed $g$ and $g'$ are in the brain, the closer their cosine
similarity is to 1.\\

 Once the co-expressions have been computed for all pairs of genes in the Allen Brain Atlas,
 they are naturally arranged in a matrix, denoted by $\CAtlas$, with the genes 
arranged in the same order as the list of genes in the atlas:
\begin{equation}
{C^{\mathrm{atlas}}}(g,g')= {\mathrm{coExpr}}(g, g') \;\;\;  1\leq g, g' \leq \GAtlas, 
\end{equation}
where $\GAtlas$ is the total number of genes included in the dataset (see 
 the next section for more details on this choice).
The matrix $\CAtlas$ is symmetric and its diagonal
entries are all equal to one. This diagonal
is trivial in the sense that it expresses the perfect 
 alignment of any vector in voxel space with itself.
 When we consider the distribution of the entries of the
 co-expression matrix, we really mean the distribution of the
 upper-diagonal coefficients.\\

Given a set of genes (with $\GSet$ elements) curated from the
literature, possibly coming from different studies, one may ask if the
brain-wide expression profiles of these genes (or a subset thereof)
are closer to each other than expected by chance, using the full atlas
as a probabilistic universe.  The set of genes for which brain-wide
expression data are available from the Allen Atlas of the adult mouse
brain consists of 4,104 genes, which is of the same order of magnitude
as the total number of genes in the mouse genome. The number of sets
of genes of a given size that can be drawn from the atlas therefore
grows quickly with the size of the set.  To study the co-expression
properties of the chosen set of genes, a $\GSet$-by-$\GSet$ matrix
$\CSet$ can be extracted from the whole co-expression matrix
$\CAtlas$. A set of strongly co-expressed genes corresponds to a
matrix $\CSet$ {\emph{with large coefficients}}. To formalise this
idea, we propose to study the matrix in terms of the underlying
graph. There are $\GAtlas!$ ways of ordering the genes in the
Atlas. They give rise to different co-expression matrices, related by
similarity transformation. But the {\emph{sets}} of highly
co-expressed genes are invariant under these transformations. The
co-expression matrix can be mapped to a weighted graph in a
straightforward way. The vertices of the graph are the genes, and the
edges are as follows:\\ 
- genes $g$ and $g'$ are linked by an edge if
their co-expression ${\mathrm{coExpr}}(g,g')$ is strictly
positive.\\ 
- If an edge exists, it has weight ${\mathrm{coExpr}}(g,g')$.\\
 
\begin{figure}
\begin{center}
\includegraphics[width=\figWidth,keepaspectratio]{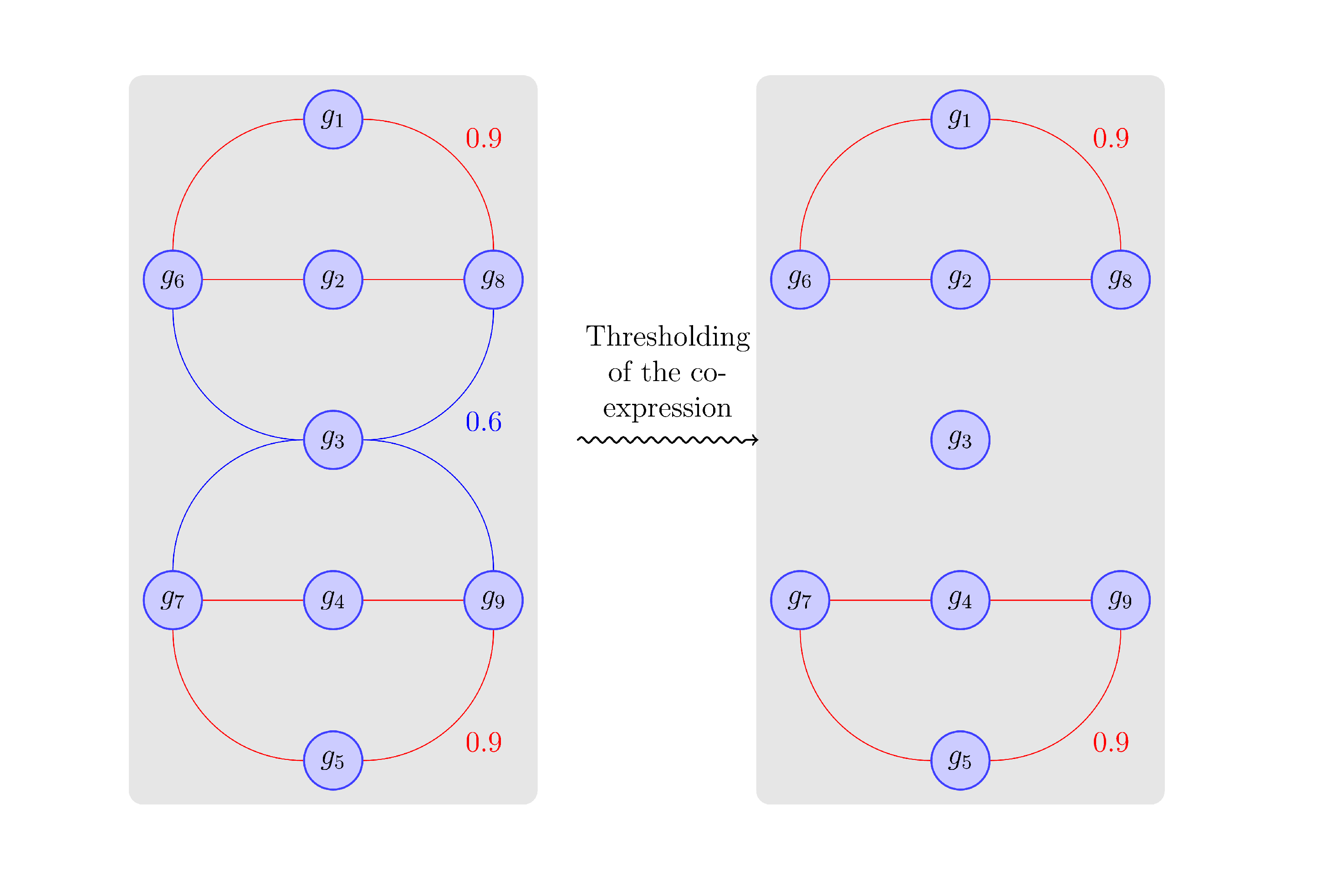}
\end{center}
\caption{A toy model with 9 genes, and only three distinct values of
  co-expression, 0, 0.6 and 0.9, for simplicity. Before any
  thresholding procedure is applied (on the left-hand side of the
  figure), there is one connected component. The average and maximum
  size of connected components are both 9. The graph on the right-hand
  side is obtained by a thresholding procedure at a threshold of
  $0.6$. There are three connected connected components, the maximal
  size is $4$, and the average size if $3$.}
\label{graphMotivation}
\end{figure}

 We have to define large co-expression matrices in relative
 terms, using thresholds on the value of co-expression that describe
 the whole set of possible values. The entries of the
 co-expression matrices are numbers between $0$ and $1$ by
 construction. We define the following thresholding procedure on
 co-expression graphs: given a threshold $\rho$ between $0$ and $1$,
 and a co-expression matrix (which can come from any
 set of genes in the Allen Atlas), put to zero all the coefficients
 that are lower than the threshold (see Figure \ref{graphMotivation} for an illustration on a toy-model with 9 genes).\\ 

 The graph corresponding to a co-expression matrix
 has connected components, and each connected component
has a certain number of genes in it.
The graph properties of $C^{\mathrm{set}}$  can be studied
 by computing the average and maximal size of the connected components 
 at every value of the threshold. This induces functions of the threshold
 that can be compared to those obtained from $\numDraws$ random sets of genes 
 of the same size $\GSet$ (these
 computations on random 
 sets of genes correspond to the two green boxes 
 in Figure \ref{flowChart}, see Supplementary Materials S2 for mathematical details).

\subsection{Cumulative distribution functions of co-expression}
To complement the graph-theoretic approach, we can study the cumulative
distribution function of the entries of the co-expression matrix of
the set of genes to study, and compare it to the one resulting from
random sets of genes of the same size (see Supplementary Materials S2
and S3 for mathematical details). For every number between 0 and 1, the
empirical cumulative distribution function of $\CSet$, denoted by
${\mathrm{CDF}}^{\mathrm{set}}$ is defined as the fraction of the
entries of the upper-diagonal part of the co-expression matrix that
are smaller than this number.\\

  To compare the co-expression network of interest $\CSet$ to
  random networks of the same size, the procedure is exactly the same
  as with the thresholded matrices, except that the quantities
  computed from the $\numDraws$ random draws are cumulative
  distribution functions rather than connected components (see
  Supplementary Materials for mathematical details). For each random 
  set of $\GSet$ genes drawn from the Allen Atlas,
   one can compute the empirical distribution
  function of the corresponding submatrix of $\CAtlas$, and
  average over the draws. The average over the draws converges towards
  the one of a typical network of $\GSet$ genes when the number 
 of random draws is sufficiently large.\\

\subsection{Comparison to classical neuroanatomy}

Given a brain region $\omega_r$, $1\leq r \leq R$, where 
 $R$ is the number of brain regions 
 in the Allen Reference Atlas \cite{AllenAtlas} (to which gene expression
 data are registered), the fitting score 
 of a brain-wide function $f$ in this region, or $\phi_r(f)$ 
can be defined \cite{markerGenes} as the cosine distance
 between this function and the characteristic function 
 of the region. It is formally the same
as the co-expression of a gene whose expression energy would 
 be the brain-wide function, and a another gene 
 that would be uniformly expressed in the region, and nowhere else:
\begin{equation}
\phi_r(f) = \sum_{v\in\Omega} \frac{f(v){\mathbf{1}}(v\in\omega_r)}
{\sqrt{\sum_{u\in\Omega} f(u)^2}\sqrt{\sum_{w\in\Omega} {\mathbf{1}}(w\in\omega_r)^2}}.
\label{fittingScore}
\end{equation}
 The distribution of fitting scores in all the brain regions
 for sets of $\GSet$ genes
   can be simulated by the Monte Carlo methods described 
 in Supplementary Materials S4.\\

 Even though clustering methods \cite{methodsPaper} have shown 
 that the correspondence between large sets of genes and brain regions
 in the Allen Atlas is not perfect, it is possible to detect
  small subsets 
 of a set of genes curated from the literature to have 
 exceptionally good fitting properties in some brain regions (see Figure
 \ref{coExprComponent2} for an example of a set of 3 genes detected to 
 fit the striatum significantly better than expected by chance).

\section{Applications}
\subsection{Choice of genes: coronal and sagittal atlases}

 The notion of an atlas of gene expression in the adult mouse brain rests 
on the assumption that there is a constant component across all brains
  at the final stage of development (the developmental atlas adresses the challenge 
 of measurement of this component at earlier stages \cite{developmentalAtlas}).
 For an account of the standardization process that began in 2001 and led to the 
 data generation and release of the Allen Brain Atlas, see \cite{AllenFiveYears}.\\

\begin{figure}
 \centering 
\includegraphics[width=\figWidth,keepaspectratio]{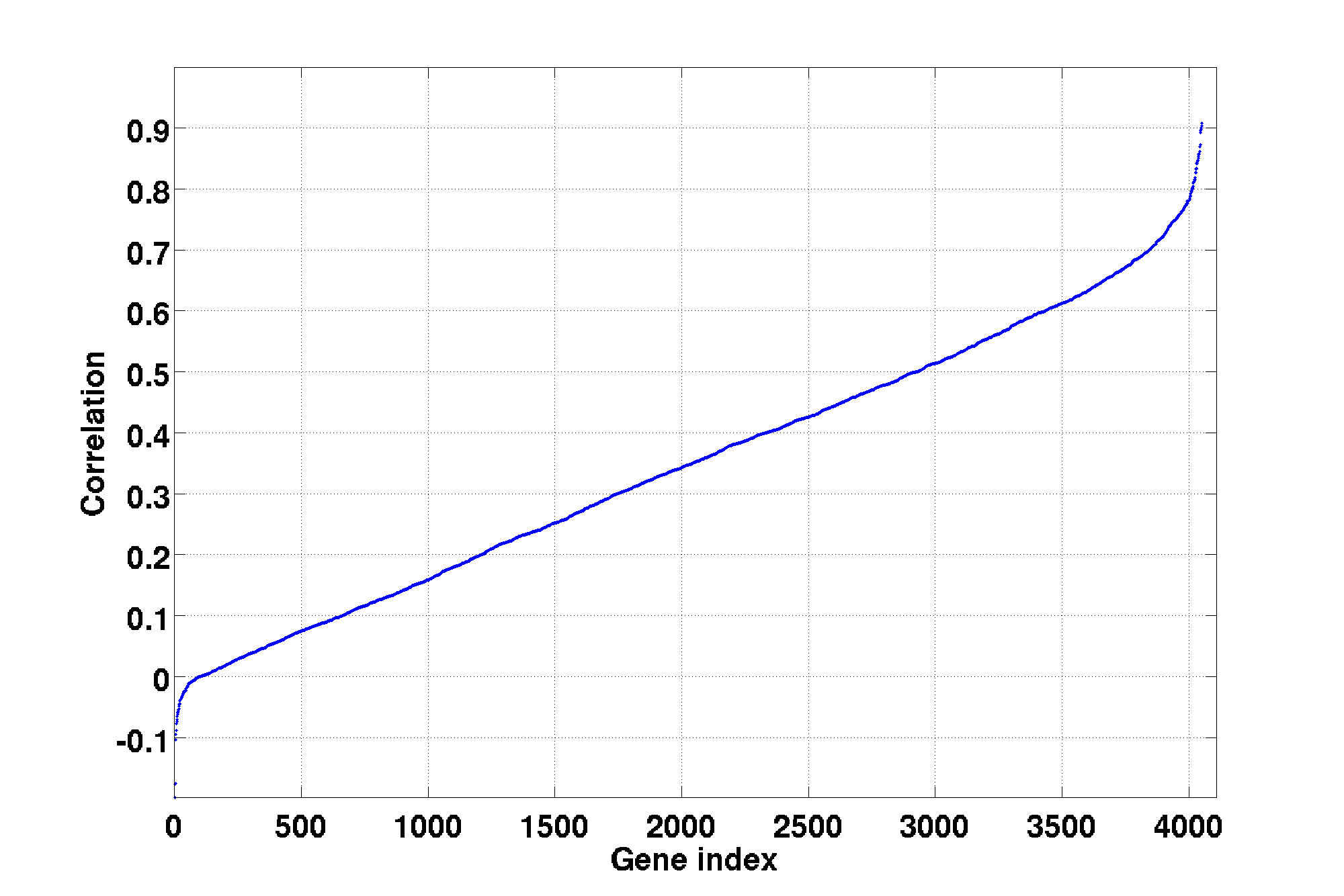}
 \caption{Sorted correlation coefficients between expression energies evaluated from sagittal and coronal sections in the 
left hemisphere of the mouse brain.}
 \label{fig:sortedCorrelations}
\end{figure}

The issue of reproducibility of ISH data can been addressed in several
ways during the analysis of data.  In NeuroBlast, the user can specify
a given image series as input.  The \toolboxName$\;$ toolbox
(\toolboxNameAcronym) is based on the analysis of the matrix of
expression energies \ref{expressionEnergy}, whose columns consist of
brain-wide gene-expression data. This restricts the choice of genes to
be analyzed in by \toolboxNameAcronym$\;$ to the 4,104 genes for which
a brain-wide, coronal atlas was developed. For these genes, sagittal,
registered data are also available in the left hemisphere. We computed
correlation coefficients between sagittal and coronal data. The
left-right correlation coefficients are not all positive.  Sagittal
datasets usually come from brain sections taken from the left
hemisphere only.  Hence the computation of correlation between
(co-registered) sagittal and coronal data has to be restricted to the
voxels belonging to the left hemisphere. For each gene $g$ in the
coronal atlas, we computed the following correlation coefficient
between sagittal and coronal data:
\begin{equation}
\rho_{\mathrm{sagittal/coronal}}( g )= \frac{\sum_{v\in{\mathrm{left\;hemisphere}}}E_{\mathrm{sagittal}}(v,g) E_{\mathrm{coronal}}(v,g)}{\sqrt{\sum_{v\in\mathrm{left\;hemisphere}}E_{\mathrm{sagittal}}(v,g)^2\sum_{v\in\mathrm{left\;hemisphere}}E_{\mathrm{coronal}}(v,g)^2}},
\label{correlation}
\end{equation}
where $E_{\mathrm{sagittal}}$ and $E_{\mathrm{coronal}}$ are the
voxel-by-gene matrices of Equation \ref{expressionEnergy} for sagittal
and coronal data respectively. The results are shown on Figure
\ref{fig:sortedCorrelations}. Some genes have negative correlation
between sagittal and coronal data. The gene with highest value of
$\rho_{\mathrm{sagittal/coronal}}$ is {\emph{Tcf7l2}}. The present
study focuses on genes for which the correlation is larger than the
25th percentile of the distribution of
$\rho_{\mathrm{sagittal/coronal}}$.\\

 This set of $GAtlas := 3,041$ genes serves as a reference set to
 which special sets of genes can be compared using the methods
 described above.  In particular, this choice excludes all the genes
 with negative correlation. Other user-defined choices of genes are
 possible within the coronal atlas. They can be implemented by
 modifying the data matrix \ref{expressionEnergy} and the list of
 genes corresponding to its columns in \toolboxNameAcronym.\\

\begin{figure}
\centering
\subfloat[]{\includegraphics[width=5.5in,keepaspectratio]{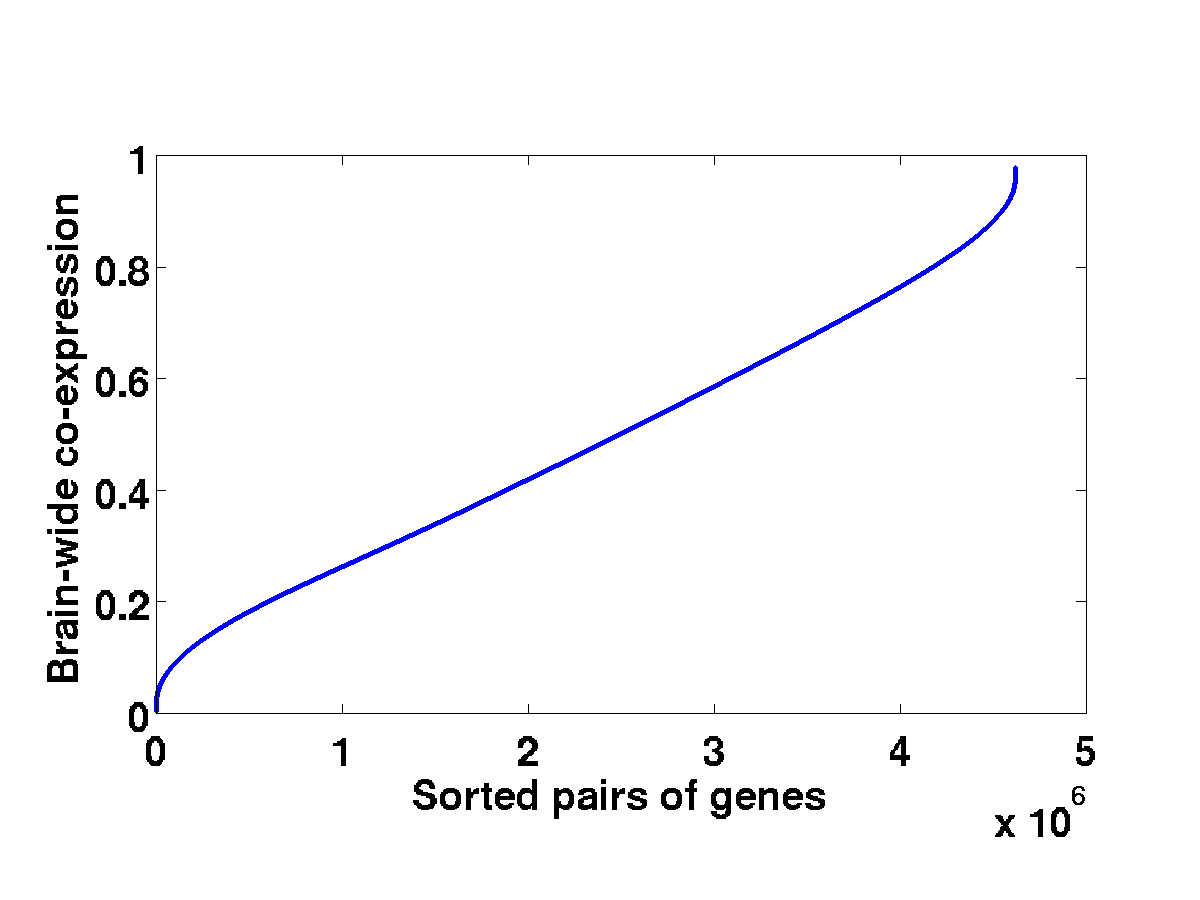}}\quad
\subfloat[]{\includegraphics[width=5in,keepaspectratio]{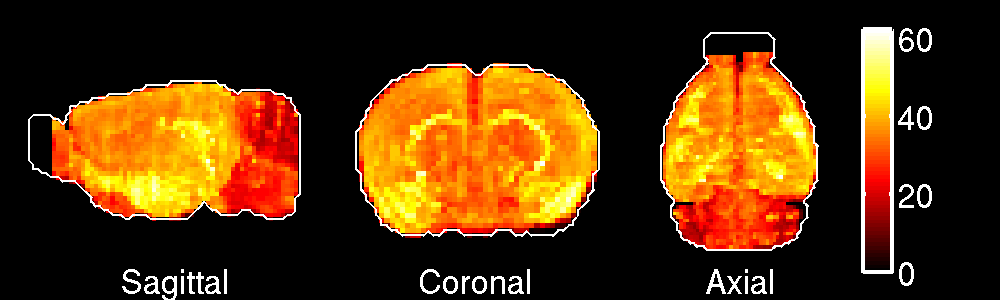}}\quad
\subfloat[]{\includegraphics[width=5in,keepaspectratio]{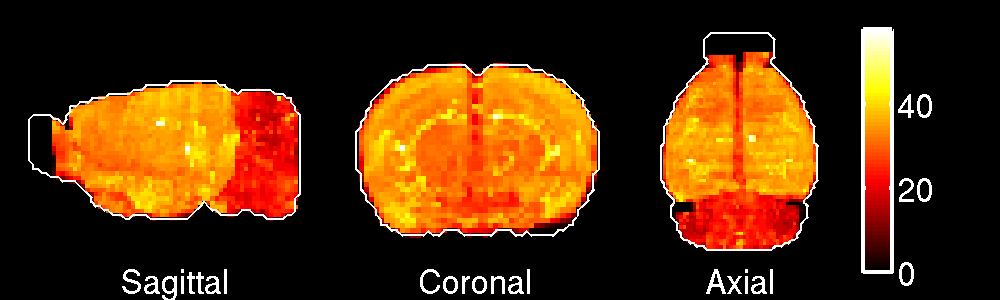}}\quad
\caption{(a) Sorted elements of the upper-diagonal part of the co-expression 
 matrix of the coronal atlas, $\CAtlas$. (b) Maximal-intensity projection of the expression energy of 
 {\emph{Atp6v0c}}. (c) Maximal-intensity projection of the expression energy of 
 {\emph{Atp2a2}}. The pair of genes ({\emph{Atp6v0c}},{\emph{Atp2a2}}) has the highest 
 co-expression in the coronal atlas, 0.9781.}
\label{coExpressionDistr} 
\end{figure}

The sorted entries of the upper-diagonal part of the induced co-expression
matrix $\CAtlas$ are plotted on Figure \ref{coExpressionDistr}(a).
The pair of genes with highest co-expression are {\emph{Atp6v0c}} and
{\emph{Atp2a2}}, whose expression profiles are plotted on Figures
\ref{coExpressionDistr}(b,c). The profile of co-expressions
 is fairly linear, except at the end of the spectrum, which motivates
 a uniform exploration of the interval $[0,1]$ when studying 
 co-expression networks (see the pseudocode in 
 Supplementary Materials S2).

\subsection{Application to a set of addiction-related genes}

\begin{figure}
\centering
\includegraphics[width=\figWidth,keepaspectratio]{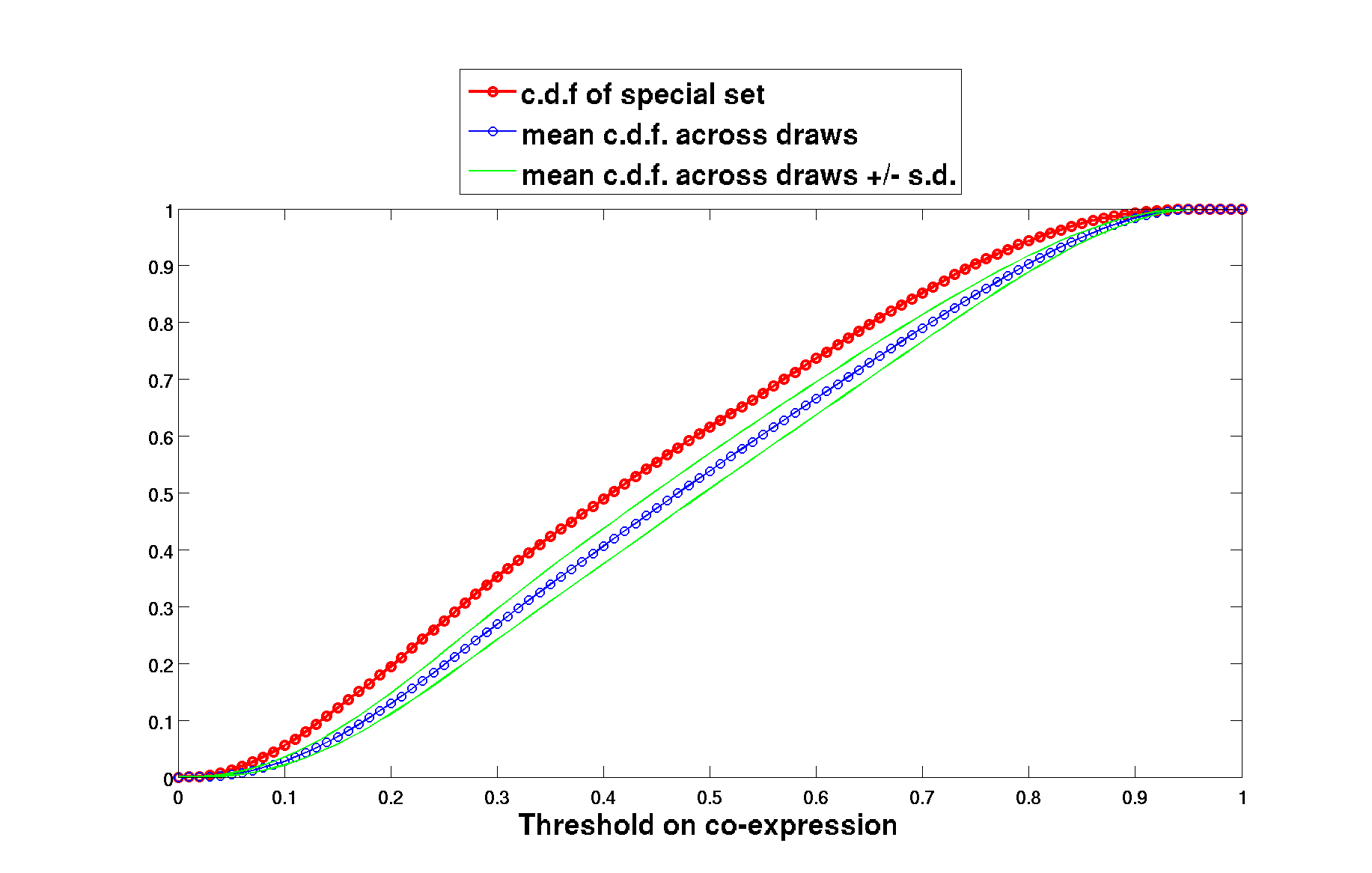}
\caption{Cumulated distribution function of the upper-diagonal entries
  of the co-expression matrix of 288 genes (the special co-expression
  network $\CSet$ of the flowchart \ref{flowChart}) from the NicSNP
  database, for which mouse orthologs are found in the Allen Atlas of
  the adult mouse brain. As the red curve (or
  $\mathrm{CDF}^{\mathrm{set}}$, Equation \ref{CDFSet}) sits above the
  simulated average of the simulated mean of CDFs (or
  $\langle\mathrm{CDF}\rangle$, Equation \ref{meanCDF}) of
  co-expression networks of 288 genes, at low values of the threshold,
  the special set as a whole appears to be less co-expressed than
  expected by chance.}
\label{cumulCoExprFunction}
\end{figure}

\begin{figure}
\centering
\includegraphics[width=\figWidth,keepaspectratio]{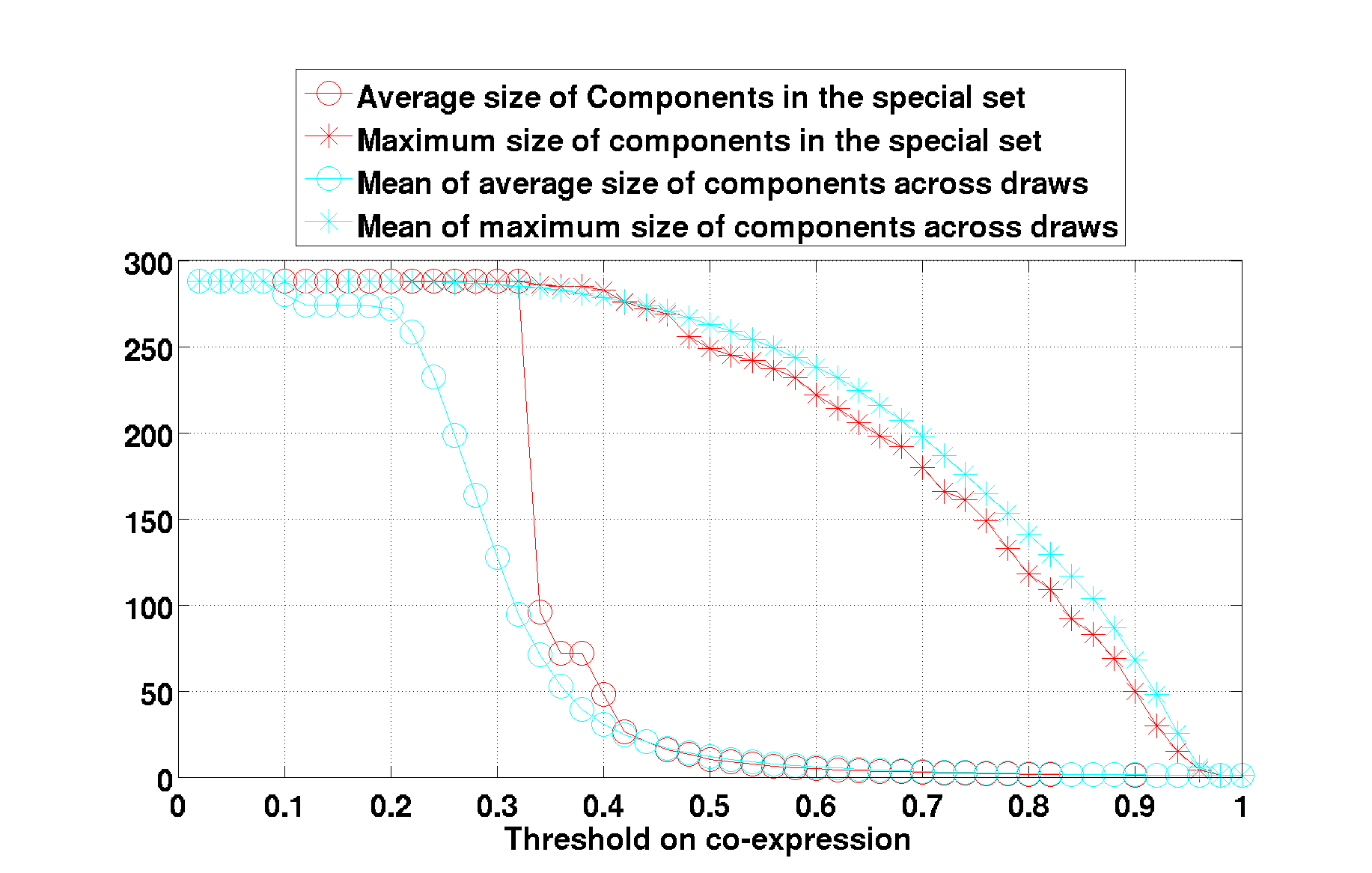}
\caption{{\bf{Monte Carlo analysis of the graph underlying the
      co-expression matrix of 288 genes from the NicSNP database.}}
  Average and maximum size of connected components as a function of
  the threshold (the quantities $\mathcal{A}(\rho)$ and
  $\mathcal{M}(\rho)$ defined in Equations \ref{averageDef} and
  \ref{maximumDef} are plotted in red circles and red stars respectively,
 the quantities $\langle\mathcal{A}(\rho)\rangle$ and
  $\langle\mathcal{M}(\rho)\rangle$ defined in Equations \ref{meanAverageDef} and
  \ref{meanMaximumDef} are plotted in cyan circles and cyan stars respectively).}
\label{coExpressionSet}
\end{figure}

\begin{figure}
\centering
\includegraphics[width=\figWidth,keepaspectratio]{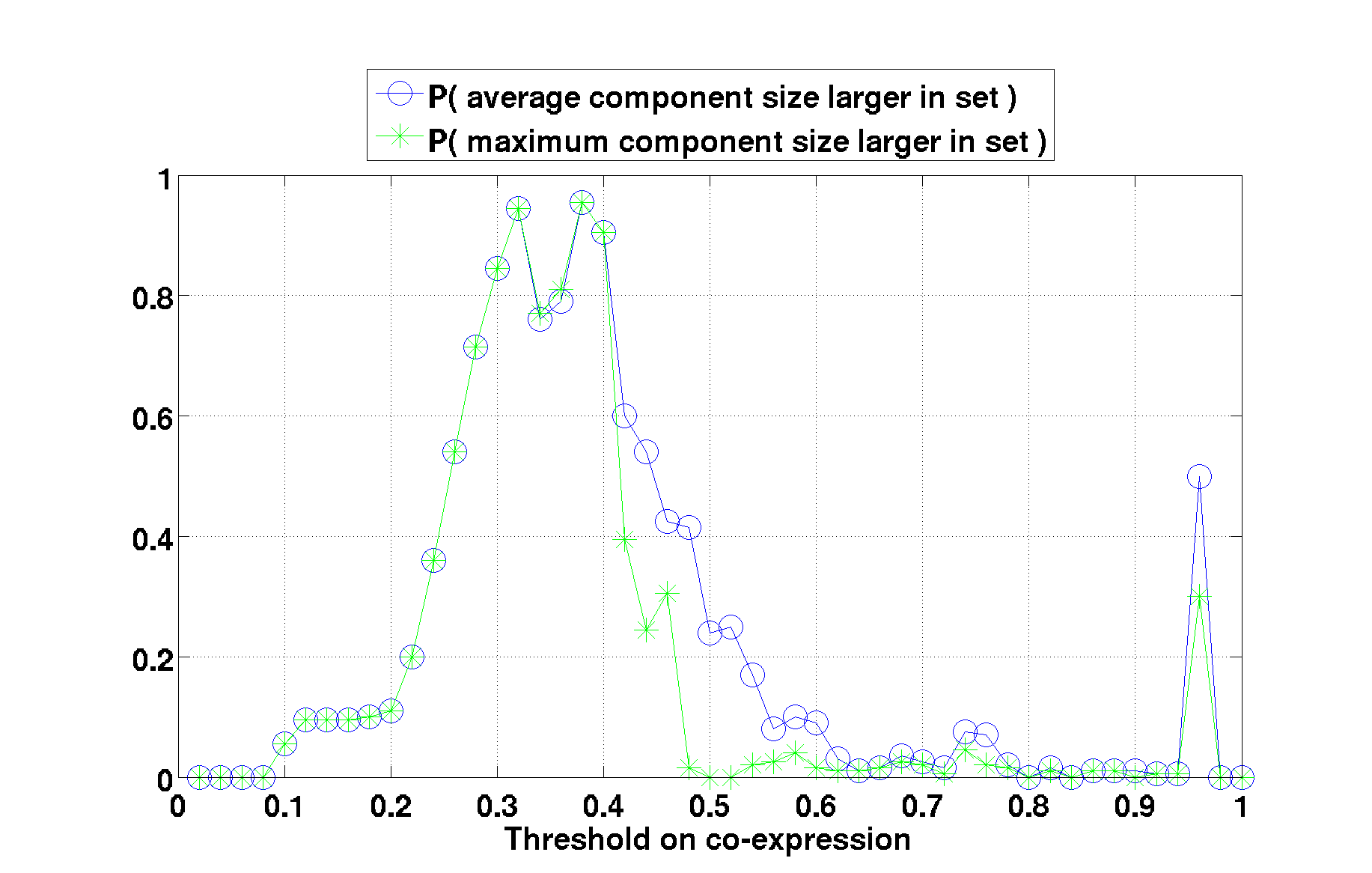}
\caption{{\bf{Monte Carlo analysis of the graph underlying the co-expression matrix of 288 genes 
 from the NicSNP database.}} Estimated probabilities for the average and maximum size of connected components to be larger than in random sets of genes of the same size
 (the quantities defined in Equations \ref{probaAvg} and \ref{probaMax} respectively).}
\label{coExpressionProba}
\end{figure}

The methods reviewed above were
  applied to a set of 288 genes related to nicotine addiction \cite{NicSNP}, retrieved from the NicSNP database\footnote{{\ttfamily{http://zork.wustl.edu/nida/Results/data1.html}}}.
 The simulation of the cumulative distribution function 
 of co-expression networks of size 288 can be compared 
 to the one of the special set, and plotted together
 on \ref{cumulCoExprFunction}. Since the CDF of the special sets
 is larger than average at low values 
 of co-expression, the special set is not more co-expressed 
 as a whole than expected by chance. This is confirmed by 
 the statistics of graph properties of networks of 288 genes (Figures
 \ref{coExpressionSet} and \ref{coExpressionProba}). See \cite{autismRelated} for a set 
 of autism-related genes that is more co-expressed
 in the brain than expected by chance).\\
\\

\clearpage

\begin{figure}
\centering
\includegraphics[width=\figWidth,keepaspectratio]{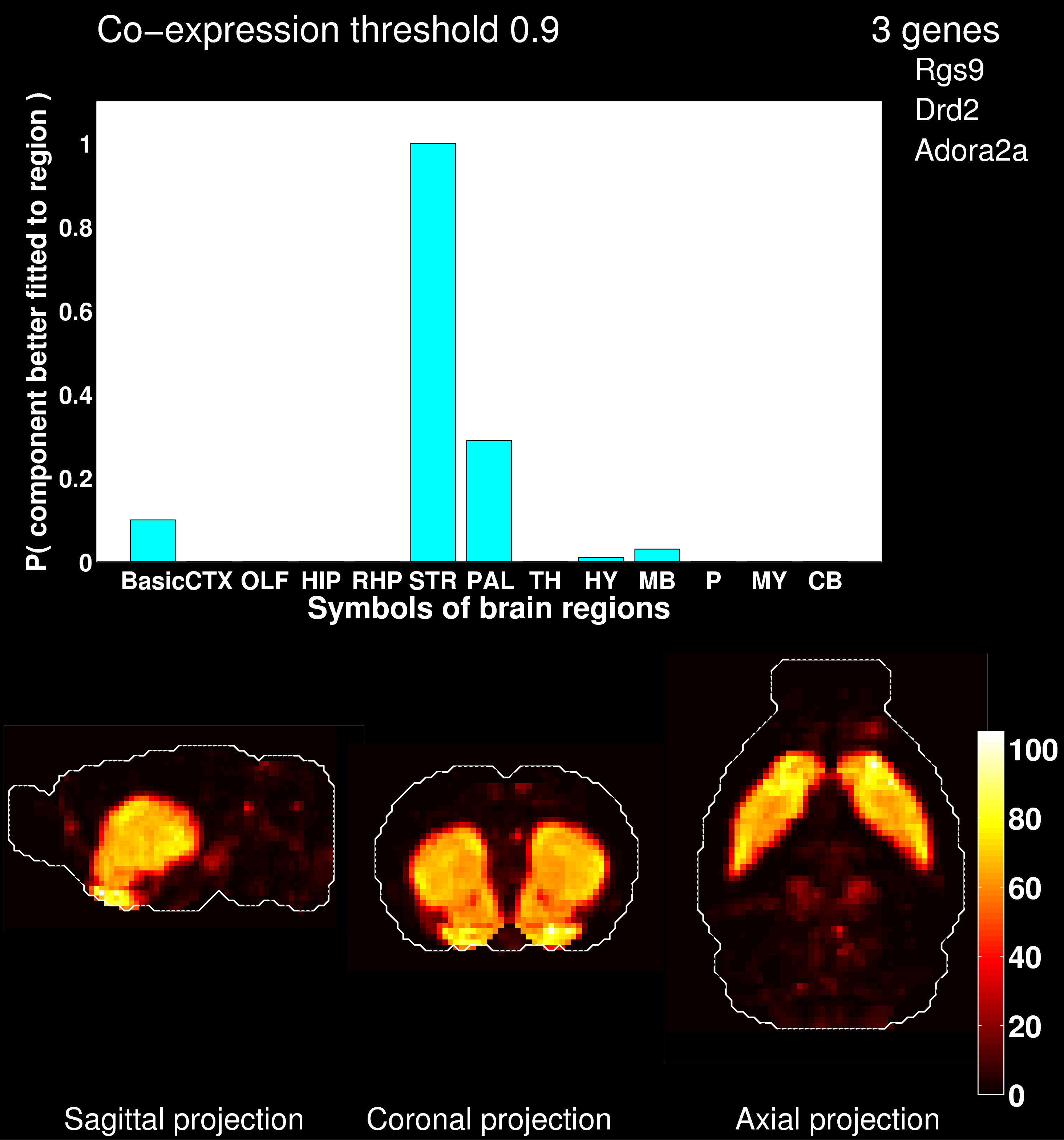}
\caption{One of the connected components of the co-expression network,
 at co-expression threshold 0.9. It is 
better-fitted to the striatum (STR) than more than 99\% of the set of
  three genes drawn from the coronal Allen Atlas of the adult mouse brain. The 
 symbols for other brain regions read as follows: Basic = 'basic cell groups',
 CTX = cortex, OLF = olfactory areas, HIP = hippocampal region, 
 RHP = retrohippocampal region, PAL = pallidum, TH = thalamus, 
 HY = hypothalamus, MB = midbrain, P = pons, MY = medulla, CB = cerebellum.}
\label{coExprComponent2}
\end{figure}


 However, the graph-based procedure returns special sets when the
 threshold on co-expression goes from 1 to 0, that may have
 exceptional neuroanatomical properties compared to sets of the same
 size, even if this does not affect the distribution of average and
 maximal size of connected components.  For each of the connected
 components, the sum of expression energies can also be compared to
 the partition(s) of the brain given by the ARA, inducing fitting
 scores in each brain regions (see Supplementary Materials S4 for
 mathematical details).  The probability for each connected component
 of thresholded co-expression networks to have a larger fitting score
 in a given brain region can be estimated.  Imposing a threshold on
 this probability (99\% for instance) returns sets of genes with
 exceptional anatomical properties.  For the coarsest partition of the
 left hemisphere, a small set of 3 genes ({\emph{Rgs9, Drd2,
     Adora2a}}) connected at a co-expression of 0.9, is in the 99th
 percentile of fitting scores in the striatum (see Figure
 \ref{coExprComponent2} for a bar diagram of the estimate of
 $P$-values of fitting scores, and a maximal-intensity projection of
 the sum of the expression energies of these genes). Even though this
 set of genes is not exceptional in terms of its size at this value of
 the co-expression threshold, it has exceptional anatomical
 properties.
 
\section{Conclusion and outlook}

 The restriction of the first release of the \toolboxName$\;$ toolbox to the
 coronal atlas of the adult mouse brain corresponds to a restriction
 to genes for which brain-wide data are available. However, the
 sagittal atlas of the adult mouse brain contains more than $20,000$
 genes, which are included in the Neuroblast and AGEA tools. 
 The second release of \toolboxNameAcronym$\;$ will include these
 genes and restrict the Allen Reference Atlas to voxels
 where all the genes have ISH data (these voxels correspond to 
 the left hemisphere of the brain). It would also be interesting to estimate
 the variability of the results under changes of probabilistic universe $\CAtlas$ 
 (by substuting the sagittal atlas to the coronal atlas, and
 by choosing different image series to construct the data matrix).\\

 Furthermore, the development of large-scale neuroscience is making
 comparable atlases available to the research community for other
 species (see \cite{humanBrainAtlas} for the Allen Atlas of the human
 brain, and \cite{MelloFinch,MelloData} for ZEBrA, the Zebra Finch
 Expression Brain Atlas), and the development of computational
 resources for the analysis of large datasets can be adapted from the
 Allen Atlas of the adult mouse brain to other atlases, allowing
 insights into evolution and into the validity of animal models
 (\cite{humanVsMouse}).\\

  Moreover, the size of voxels in the Allen Brain Atlas is large in
  scale of brain cells, and each voxel may contain cells of different
  types. Several studies
  \cite{RossnerCells,CahoyCells,DoyleCells,ChungCells,ArlottaCells,HeimanCells,foreBrainTaxonomy}
  have obtained cell-type-specific transcriptional profiles using
  microarray experiments.  Comparison between ISH and microarray data
  is an ongoing challenge \cite{LeeQuantitative}, and steps were taken
  in \cite{AGEACellTypes} to estimate the brain-wide density profiles
  of cell types by combining the Allen Atlas to the transcriptional
  profiles of cell types. This sheds light on 
 the cellular origin of co-expression brain-wide co-expression patterns
 of genes. The corresponding Matlab code will be
  included in the second release of \toolboxNameAcronym.

\section{Acknowledgments}
 We thank Sharmila Banerjee-Basu, Idan Menashe, Eric C. Larsen, Hemant Bokil
 and Jason W. Bohland for discussions and collaboration.  
This research is supported by the NIH-NIDA Grant 1R21DA027644-01,
{\emph{Computational analysis of co-expression networks in the mouse
    and human brain}}.

\clearpage
\section{Supplementary Materials}
\subsection{S1: Co-expression networks, graph properties}

Consider a set of genes of size $\GSet$, as in the ellipse 
 on the left-hand-side of the flowchart \ref{flowChart}.  They correspond to indices
$(g_1,\dots, g_{\GSet})$ in the columns of the voxel-by-gene matrix $E$ of
expression energies. We can construct the co-expression matrix by
extracting the coefficients of the co-expression matrix of the atlas
corresponding to these genes. Let us denote this matrix by $\CSet$:
\begin{equation}
 \CSet(i,j)=\CSet(g_i,g_j).
\label{extractionEquation}
\end{equation}

 After applying the thresholding procedure, the co-expression
matrix is mapped to a matrix $\CSet_\rho$:
\begin{equation}
 \CSet_\rho(i,j)=\CSet(i,j) \times {\mathbf{1}}(\CSet(i,j)\geq \rho).
\label{thresholdEquation}
\end{equation}
  Then for every integer $k$ between 1 and $\GSet$ we can count the number
  $N_\rho(k)$ of connected components of $C_\rho$ that have exactly
  $k$ genes in them (using Tarjan's algorithm \cite{Tarjan},
  implemented as the function {\ttfamily{graphconncomp.m}} in Matlab).
  We can study the average size of connected components of thresholded
  co-expression networks and the size of the largest connected component:
\begin{equation}
\mathcal{A}( \rho ) = \frac{\sum_{k = 1}^{\GSet} k  N_\rho(k)}{\sum_{k = 1}^G  N_\rho(k) },
\label{averageDef}
\end{equation}
\begin{equation}
\mathcal{M}( \rho ) = \mathrm{max}\left\{k \in [1..\GSet],  N_\rho(k) > 0 \right\},
\label{maximumDef}
\end{equation}
as a function of the threshold $\rho$.
$\mathcal{A}( 0 )$ and $\mathcal{M}( 0 )$ 
 both equal the size of the set of genes, as the whole set is connected
 before any thresholding procedure is applied.
  At large thresholds every single singe is disconnected from the 
other genes, as having co-expression equal to one is equivalent to having exactly the same expression across the whole brain. 
So at threshold 1 all the connected components have size one, 
and $\mathcal{A}( 1 ) = \mathcal{M}( 1 )= 1$.\\

\subsection{S2: Monte Carlo study of gene networks}

We would
like to compare the properties of the matrix $C^{\mathrm{set}}$ to the
ones of $C^{\mathrm{atlas}}$. In order to eliminate the sample-size bias,
 we are going to study some properties
 of the graph underlying $C^{\mathrm{set}}$, and to compare them to the
properties of the graphs underlying submatrices of
$C^{\mathrm{atlas}}$ of the same size, $\GSet$.\\

To explore the graph property of the gene network, we have to choose a
discrete set of thresholds regularly spaced between 0 and 1,
 and to apply the procedure of Equation \ref{thresholdEquation} usuing each of these thresholds. Call
$\numDraws$ the number of random sets of genes to be drawn. The
computations can be described as follows in pseudocode:\\ 
{\ttfamily{
    1. Choose a number of thresholds $T$ to study.\\
 2. Choose a
    number of draws $\numDraws$ to be performed for each value\\
 of the threshold;\\
 3. For each integer $t$ between $1$ and
    $T$:\\ 
3.a. consider the threshold $\rho_t =
    \frac{t}{T};$
\\ 3.b. compute the connected components of the
    thresholded matrix $C^{\mathrm{set}}_{\rho_t}$, as defined in Equation \ref{thresholdEquation};
    call
    $\mathcal{M}^{\mathrm{set}}( \rho_t )$ the size of the largest
    connected component, and $\mathcal{A}^{\mathrm{r}}( \rho_t )$
     the average size of connected components;\\ 4. for each integer $n$ between $1$ and
    $\mathcal{N}$:\\ draw a random set of distinct indices of size $\GSet$
    from [1..G],\\ extract the corresponding submatrix of
    $C^{\mathrm{atlas}}$;\\ call it $C^{n}$, and repeat step 3 after
    substituting $C^{n}$ to $C^{set}$;\\ call
    $\mathcal{M}^{\mathrm{n}}( \rho_t )$ the size of the largest
    connected component of $C^{\mathrm{n}}_{\rho_t}$, and $\mathcal{A}^{\mathrm{n}}( \rho_t )$
    the average size of connected components.}}\\

At each value $\rho$ of the threshold, we therefore have:\\ - the size
of the maximal connected component $\mathcal{M}^{\mathrm{set}}( \rho
)$,\\ - a distribution of $\mathcal{N}$ numbers, each of wchich is the
size of the largest connected components of a random submatrix of the
same size as the set of genes to study, thresholded at $\rho$.\\
 
When the number of random draws is sufficiently large,
 we can estimate the means of the average and maximum sizes of 
 connected components: 
\begin{equation}
 \langle \mathcal{M}( \rho)\rangle = \frac{1}{\numDraws} \sum_{n = 1}^\numDraws \mathcal{M}^{\mathrm{n}}( \rho ),
\label{meanMaximumDef}
\end{equation}
\begin{equation}
 \langle \mathcal{A}( \rho)\rangle = \frac{1}{\numDraws} \sum_{n = 1}^\numDraws \mathcal{A}^{\mathrm{n}}( \rho ),
\label{meanAverageDef}
\end{equation}
 We
can study where in $\mathcal{M}^{\mathrm{set}}( \rho )$
(resp. $\mathcal{A}^{\mathrm{set}}( \rho )$) sits in the distribution
and estimate the probabilities of $\mathcal{M}^{\mathrm{set}}( \rho )$ 
 and $\mathcal{A}^{\mathrm{set}}( \rho )$ being larger than expected by chance:
\begin{equation}
P_{\mathrm{larger}}( \mathcal{A}^{\mathrm{set}}(\rho)) = \frac{1}{\numDraws} 
               \sum_{n=1}^\numDraws{\mathbf{1}}\left(\mathcal{A}^{\mathrm{set}}(\rho)\geq
                 \mathcal{A}^{\mathrm{n}}( \rho)\right).
\label{probaAvg}
\end{equation}
\begin{equation}
P_{\mathrm{larger}}( \mathcal{M}^{\mathrm{set}}(\rho)) = \frac{1}{\numDraws} 
               \sum_{n=1}^\numDraws{\mathbf{1}}\left( \mathcal{M}^{\mathrm{set}}(\rho)\geq
                 \mathcal{M}^{\mathrm{n}}( \rho)\right),
\label{probaMax}
\end{equation}

\subsection{S3: Cumulative distribution functions (CDFs)}

Given the $\GSet$-by-$\GSet$ co-expression matrix $\CSet$,
 consider the coefficients above the diagonal (which
are the meaningful quantities by construction) and arrange them
into a vector $C_{\mathrm{vec}}$ with $N=\GSet(\GSet-1)/2$ components:
$C_{{\mathrm{vec}}} = \{ C_{gh} \}_{1 \leq g \leq \GSet, h > g}$. The
components of this vector are numbers between 0 and 1. For every number between 0 and
1, the cumulative distribution function of $C$, denoted by
${\mathrm{CDF}}^{\mathrm{set}}$ is defined as the fraction of the components of
$C_{{\mathrm{vec}}}$ that are smaller than this number:

\begin{align}
{\mathrm{CDF}}^{\mathrm{set}}& : [ 0,1 ] \rightarrow [ 0,1 ]\\
 x& \mapsto \frac{1}{N} \sum_{k = 1}^{N} \delta_{C_{{\mathrm{vec}}}( k ) \leq x }.
\label{CDFSet}
\end{align}

For any set of genes, ${\mathrm{CDF}}^{\mathrm{set}}$ is a growing 
 function ${\mathrm{CDF}}^{\mathrm{special}}(0) = 0$ and
  ${\mathrm{CDF}}^{\mathrm{special}}(1) = 1$.
  For highly co-expressed genes, the growth of ${\mathrm{CDF}}^{\mathrm{special}}$ 
  is concentrated at high values of the argument (in a situation where all the genes 
 in the special set have the same brain-wide expression vector, all the entries 
 of the co-expression matrix equal $1$). To 
  compare the function ${\mathrm{CDF}}^{\mathrm{special}}$ to what could be 
 expected by chance, let us draw $\mathcal{N}$ random sets of $G_{\mathrm{special}}$ genes 
 from the Atlas, compute their co-expression network
 by extracting the corresponding entries from the full co-expression 
 matrix of the atlas ($C^{\mathrm{atlas}}$). This induces
 a family of $\mathcal{N}$ growing functions ${\mathrm{CDF}}_{i}, 1\leq i\leq R$
 on the interval $[0,1]$:

\begin{equation}
 \forall 1\leq i \leq \numDraws,\;\;\;\;\;{\mathrm{CDF}}_{i}: [0,1] \rightarrow [0,1].
\end{equation}

From this family of functions, we can estimate a mean cumulative
 distribution function $\langle{\mathrm{CDF}}\rangle$ of the co-expression of sets of 
 $G_{\mathrm{special}}$ genes drawn from the Allen Atlas, by taking 
 the mean of the values of ${\mathrm{CDF}}_{i}$ across the 
  random draws:\\

\begin{equation}
\forall x \in [ 0,1 ],\;\;\; \langle{\mathrm{CDF}}\rangle(x) = \frac{1}{{\mathcal{N}}}
\sum_{i=1}^{\mathcal{N}}{\mathrm{CDF}}_{i}(x).
\label{meanCDF}
\end{equation}
Standard deviations $\mathrm{CDF}^{\mathrm{dev}}$ of the distribution of CDFs are estimated 
 as follows on the interval $[0,1]$:\\

\begin{equation}
\forall x \in [ 0,1], \;\;\;\mathrm{CDF}^{\mathrm{dev}}(x) = 
\sqrt{\frac{1}{{\mathcal{N}}}\sum_{i=1}^{\numDraws}\left({\mathrm{CDF}}_{i}(x)-\langle{\mathrm{CDF}}\rangle(x)\right)^2}.
\end{equation}

\subsection{S4: Comparison to classical neuroanatomy}

Consider a system of annotation in the 
voxelized version of the ARA. Let $\Omega$
 be the set of voxels in the annotation,
 let $R$ be the total number of regions in the 
annotation, and let $\{\omega_1,\dots,\omega_R\}$
 be the regions in the annotation.
 For the sake of simplicity, the present paper 
 focusses on the coarsest annotation,
  for which $\Omega$ is the left hemisphere,
 and $R$ = 13.\\

 For a set of $\GSet$ genes, labelled $\{g_1,\dots, g_{\GSet} \}$
 in the atlas, the total brain-wide expression 
 energy is
 \begin{equation}
 E^{\mathrm{set}}(v,\{g_1,\dots, g_{\GSet} \}) = \sum_{k=1}^{\GSet} E( v, g_k).
 \label{expressionSet}
 \end{equation}
  
 Using the same Monte Carlo procedure as in 
 supplementary S2, we draw $\numDraws$ sets of genes
 from the atlas, and compute the total gene-expression
 energy defined by Equation \ref{expressionSet}
 for each of these sets. Hence for each random draw (labelled by an integer $n$ in $[1..\numDraws]$)
 one has a set of genes labelled $\{g^{n}_1,\dots, g^{n}_{\GSet} \}$,
 and the corresponding brain-wide sums of expression energies
\begin{equation}
 E^{\mathrm{rand},\mathrm{n}}(v,\{g^{n}_1,\dots, g^{n}_{\GSet}  \}) = \sum_{k=1}^{\GSet} E( v, g^{n}_k).
 \label{expressionSet}
 \end{equation}
 
 The fitting scores to each region in the ARA
can be computed both for $E^{\mathrm{set}}$ and for the 
 each of the random sets of $\GSet$ genes:
\begin{align}
\phi_r^{\mathrm{set}} &:= \phi_r( E^{\mathrm{set}} ),\\
\phi_r^{\mathrm{rand},\mathrm{n}} &:= \phi_r(E^{\mathrm{rand},\mathrm{n}}).
\label{fittingScoreDistr}
\end{align}
Hence, one can estimate the position of 
 the fitting score $\phi_r^{\mathrm{set}}$ in the distribution of fitting scores
by evaluating the folllowing fraction:
 \begin{equation}
 P^{\mathrm{set}}_r = \frac{1}{\numDraws}\sum_{n=1}^{\numDraws}{\mathbf{1}}\left(\phi_r^{\mathrm{set}}
    \geq\phi_r^{\mathrm{rand},\mathrm{n}}\right)
 \label{fitProbaEst}
 \end{equation}
 which goes to the probability for $\phi_r^{\mathrm{set}}$ being larger than expected by chance
 for a sum of $\GSet$ expression energies. 
 A histogram of \ref{fitProbaEst} is plotted on Figure \ref{coExprComponent2},
 for the regions of the coarsest annotations
 of the left hemisphere, and for a set 
 consisting of {\emph{Rgs2}}, {\emph{Drd2}} and {\emph{Adora2a}}.


\clearpage


\begin{thebibliography}{9}


\bibitem{GeneToBrain} M. Bota, H.-W. Dong and L.W. Swanson, {\emph{From
    gene networks to brain networks}}, Nature neuroscience (2003)
  {\bf{6}} (8), 795--9.
         
\bibitem{AllenGenome}
 E.S. Lein, M. Hawrylycz, N. Ao, M. Ayres, A. Bensinger, A. Bernard, A.F. Boe,
M.S. Boguski, K.S. Brockway, E.J. Byrnes, L. Chen, L. Chen, T.M. Chen, M.C. Chin, 
J. Chong, B.E. Crook, A. Czaplinska, C.N. Dang, S. Datta, N.R. Dee, {\emph{et al.}},
 {\emph{Genome-wide atlas of gene expression in the adult mouse brain.}}
Nature {\bf{445}}, 168 –176 (2007).




\bibitem{BrainAtlasInsights} S.M. Sunkin and J.G. Hohmann,
  {\emph{Insights from spatially mapped gene expression in the mouse
      brain}}, Human Molecular Genetics, 2007, Vol. 16, Review Issue 2.


\bibitem{images}  L. Ng, M. Hawrylycz, D. Haynor, {\emph{Automated high-throughput registration for localizing 3D mouse brain gene expression using ITK}}, Insight-Journal (2005).


\bibitem{neufoAllen} L. Ng, S.D. Pathak, C. Kuan, C. Lau, H. Dong, A. Sodt, C. Dang,
 B. Avants, P. Yushkevich, J.C. Gee, D. Haynor, E. Lein, A. Jones and M. Hawrylycz,
{\emph{Neuroinformatics for genome-wide 3D gene expression mapping in the mouse brain}},
IEEE/ACM Trans. Comput. Biol. Bioinform. (2007), Jul-Sep {\bf{4(3)}} 382--93.


\bibitem{AllenFiveYears} A.R. Jones, C.C. Overly and S.M. Sunkin,
  {\emph{The Allen Brain Atlas: 5 years and beyond}}, Nature Reviews
  (Neuroscience), Volume {\bf{10}} (November 2009), {\bf{1}}.


\bibitem{corrStructureAllen} M. Hawrylycz, L. Ng, D. Page, J. Morris, C. Lau, S. Faber, V. Faber, S. Sunkin,
 V. Menon, E.S. Lein, A. Jones, {\emph{Multi-scale correlation structure of gene expression in the brain}}, Neural Networks {\bf{24}} (2011) 933--942.



\bibitem{addiction} Computational analysis of user-defined sets of 
genes from the Allen Atlas of mouse and human brain can be
 conducted online at {\ttfamily{addiction.brainarchitecture.org}} 

\bibitem{toolboxManual} P. Grange, J.W. Bohland, M. Hawrylycz and 
 P.P. Mitra, {\emph{Brain Gene Expression Analysis: a MATLAB toolbox for the analysis of brain-wide gene-expression data}}, {\ttfamily{arXiv:1211.6177 [q-bio.QM]}}, code downloadable at \toolboxLink. 



\bibitem{KARG} C.Y. Li, X. Mao, L. Wei (2008) Genes and (common)
  pathways underlying drug addiction. PLoS Comput Biol 4(1):e2. doi:10.1371/journal.pcbi.0040002. Data can be retrieved from {\ttfamily{http://karg.cbi.pku.edu.cn/}}
 


\bibitem{NicSNP} S.F. Saccone, N.L. Saccone, G.E. Swan, P.A.F. Madden,
  A.M. Goate, J.P. Rice and L.J. Bierut, {\emph{Systematic biological
      prioritization after a genome-wide association study: an
      application to nicotine dependence}}, Bioinformatics (2008),
  {\bf{24}}, 1805--1811.



\bibitem{BLAST} S.F. Altschul, W. Gish, W. Miller, E.W. Myers and D.J. Lipman, {\emph{Basic local alignment search tool}},
 J. Mol. Biol. {\bf{215}}, 403--410 (1990).

\bibitem{NeuroBlast} L. Ng {\emph{et al.}}, {\emph{NeuroBlast: a 3D spatial homology search tool for gene
 expression}}, BMC Neuroscience 2007, {\bf{8}}(Suppl 2):P11.


\bibitem{AGEA} L. Ng, A. Bernard, C. Lau, C.C. Overly, H.-W. Dong,
  C. Kuan, S. Pathak, S.M. Sunkin, C. Dang, J.W. Bohland, H. Bokil,
  P.P. Mitra, L. Puelles, J. Hohmann, D.J. Anderson, E.S. Lein,
  A.R. Jones, M. Hawrylycz, {\emph{An anatomic gene expression atlas
      of the adult mouse brain}}, Nature Neuroscience {\bf{12}}, 356 -
  362 (2009).



\bibitem{ZhangFramework} B. Zhang and S. Horvath, {\emph{A general
    framework for weighted gene co-expression network analysis}}.

\bibitem{feeding} P.K. Olszewski, J. Cederna F. Olsson, A.S. Levine and H.B. Schioth,
{\emph{Analysis of the network of feeding neuroregulators using the Allen Brain Atlas}}, Neurosci. Biobehav. Rev. {\bf{32}}, 945--956 (2008). 



\bibitem{AllenAtlas} H.-W. Dong,
\emph{The Allen reference atlas: a digital brain atlas of the C57BL/6J male mouse}, 
	  Wiley, 2007.


\bibitem{markerGenes} P. Grange and P.P. Mitra, {\emph{Computational neuroanatomy and gene expression: Optimal sets of marker genes for brain regions}}, IEEE, in CISS 2012, 46th annual conference on Information Science and Systems (Princeton), {\ttfamily{arXiv:1205.2721 [q-bio.QM]}}.




\bibitem{methodsPaper}
J.W. Bohland, H. Bokil, C.-K. Lee, L. Ng, C. Lau, C. Kuan, M. Hawrylycz and P.P. Mitra,
{\emph{Clustering of spatial gene expression patterns in the mouse brain and comparison with classical neuroanatomy}},
Methods, Volume {\bf{50}}, Issue 2, February 2010, Pages 105-112.


\bibitem{LauExploration} C. Lau, L. Ng, C. Thompson, S. Pathak,
  L. Kuan, A. Jones and M. Hawrylycz, {\emph{Exploration and
      visualization of gene expression with neuroanatomy in the adult
      mouse brain}}, BMC Bioinformatics 2008, {\bf{8}}:153.


\bibitem{digitalAtlasing} M. Hawrylycz, R.A. Baldock, A. Burger,
  T. Hashikawa, G.A. Johnson, M. Martone, L. Ng, C. Lau, S.D. Larsen,
  J. Nissanov, L. Puelles, S. Ruffins, F. Verbeek, I. Zaslavsky1 and
  J. Boline, {\emph{Digital Atlasing and Standardization in the Mouse
      Brain}}, PLoS Computational Biology {\bf{7}} (2) (2011).

\bibitem{AllenWebServiceSite} The Allen Brain Atlas can be used online at
  {\ttfamily{www.brain-map.org/}}.


\bibitem{developmentalAtlas} The developmental atlas of the mouse brain 
 is available from {\ttfamily{http://developingmouse.brain-map.org/}}


\bibitem{autismRelated} I. Menashe, P. Grange, E.C. Larsen,
  S. Banerjee-Basu and P.P. Mitra, {\emph{Co-expression profiling of autism genes in the mouse brain}}, SFN Abstracts 2012, and in preparation.
 

\bibitem{humanBrainAtlas} M. Hawrylycz {\emph{et al.}}, {\emph{An anatomically comprehensive atlas of the adult human brain transcriptome}}, Nature {\bf{489}}, 391–399.


\bibitem{MelloFinch} W.C. Warren, D.F. Clayton, H. Ellegren, A.P. Arnold,
L.W. Hillier, A. Kunstner, S. Searle, S. White, A.J. Vilella and
 S. Fairley {\emph{et al.}} (2010), {\emph{The genome of a songbird}}. Nature, 464,
757–762.

\bibitem{MelloData} Data can be retrieved from the ZEBrA database.
(Oregon Health and Science University, Portland, OR 97239;
 {\ttfamily{http://www.zebrafinchatlas.org}}).


\bibitem{humanVsMouse} J.A. Miller, S. Horvath and D.H. Geschwind,
 {\emph{Divergence of human and mouse brain transcriptome
 highlights Alzheimer disease pathways}}, Proc. Natl. Acad. Sci. U.S.A. (2010) {bf{107(28)}}:12698-703.


\bibitem{foreBrainTaxonomy} K. Sugino, C.M. Hempel, M.N. Miller, A.M. Hattox, P. Shapiro, C. Wu, Z.J. Huang and
 S.B. Nelson, {\emph{Molecular taxonomy of major neuronal classes in the adult mouse forebrain}},
 Nature Neuroscience {\bf{9}}, 99-107 (2005). 

\bibitem{ChungCells} C.Y. Chung, H. Seo, K.C. Sonntag, A. Brooks, L. Lin and
 O. Isacson, {\emph{Cell-type-specific gene expression of midbrain dopaminergic neurons
reveals molecules involved in their vulnerability and protection}}. Hum. Mol. Genet. (2005)
{\bf{14}}: 1709--1725.


\bibitem{ArlottaCells} P. Arlotta, B.J. Molyneaux, J. Chen, J. Inoue,
 R. Kominami {\emph{et al.}} (2005) {\emph{Neuronal subtype-specific genes
 that control corticospinal motor neuron development in vivo}},
 Neuron {\bf{45}}: 207--221.


\bibitem{HeimanCells} M. Heiman, A. Schaefer, S. Gong, J.D. Peterson,
  M. Day, K.E. Ramsey, M. Suárez-Fari$\tilde{\mathrm{n}}$as, C. Schwarz, D.A. Stephan,
  D.J. Surmeier, P. Greengard and N. Heintz (2008) {\emph{A
      translational profiling approach for the molecular
      characterization of CNS cell types}}, Cell {\bf{135}}:
  738--748.


\bibitem{RossnerCells} M.J. Rossner, J. Hirrlinger, S.P. Wichert, C. Boehm, D. Newrzella, H. Hiemisch, G. Eisenhardt, C. Stuenkel, O. von Ahsen and K.A. Nave, {\emph{Global transcriptome analysis of genetically identified neurons in the adult cortex}},
J. Neurosci. 2006 {\bf{26(39)}} 9956-66.

\bibitem{CahoyCells} J.D. Cahoy, B. Emery, A. Kaushal, L.C. Foo,
  J.L. Zamanian, K.S. Christopherson, Y. Xing, 
  J.L. Lubischer, P.A. Krieg,
  S.A. Krupenko, W.J. Thompson WJ and B.A. Barres, {\emph{A transcriptome database
      for astrocytes, neurons, and oligodendrocytes: a new resource
      for understanding brain development and function}},
  J. Neurosci. 2008 {\bf{28(1)}} 264-78.
 

\bibitem{DoyleCells} J.P. Doyle, J.D. Dougherty, M. Heiman, E.F. Schmidt, T.R. Stevens, G. Ma, S. Bupp,
 P. Shrestha, R.D. Shah, M.L. Doughty, S. Gong, P. Greengard and N. Heintz, 
{\emph{Application of a translational profiling approach for the comparative
analysis of CNS cell types}},
Cell (2008) {\bf{135(4)}} 749-62. 


\bibitem{LeeQuantitative} C.K. Lee {\emph{et al.}}, {\emph{Quantitative methods for genome-scale analysis
 of in situ hybridization and correlation with microarray data}}, Genome Biol. {\bf{9}}, R23 (2008).


\bibitem{AGEACellTypes} P. Grange, J. Bohland, H. Bokil, S. Nelson,
  B. Okaty, K. Sugino, L. Ng, M. Hawrylycz and P.P. Mitra, {\emph{A
      cell-type based model explaining co-expression patterns of genes
      in the brain}}, {\ttfamily{arXiv:1111.6217 [q-bio.QM]}}.


\bibitem{Tarjan} R. E. Tarjan, {\emph{Depth first search and linear
    graph algorithms}}, SIAM Journal on Computing, {\bf{1(2)}}:146-160, 1972.







\end{thebibliography}
\end{document}